\documentclass[10pt, conference, compsocconf]{IEEEtran}
\ifCLASSINFOpdf
\else
\fi

\usepackage{cite}
\usepackage{amsmath,amssymb,amsfonts}
\usepackage{algorithmic}
\usepackage{graphicx}
\usepackage{textcomp}
\usepackage{xcolor}
\usepackage{makecell}
\usepackage{multirow}
\usepackage{booktabs}
\usepackage{etoolbox}
\usepackage[linesnumbered,ruled]{algorithm2e}

\let\labelindent\relax
\usepackage{enumitem,kantlipsum}
\usepackage{multicol}
\usepackage{tabularx}
\usepackage{subcaption}
\usepackage{bigstrut}

\newcommand\scalemath[2]{\scalebox{#1}{\mbox{\ensuremath{\displaystyle #2}}}}

\begin{document}
\title{Multi-Graph Convolution Collaborative Filtering}



% This work is done when Chen Ma worked as intern at Noah’s Ark Research Lab, Huawei
\author{\IEEEauthorblockN{Jianing Sun\IEEEauthorrefmark{1},
Yingxue Zhang\IEEEauthorrefmark{1},
Chen Ma\IEEEauthorrefmark{2}\IEEEauthorrefmark{1}, Mark Coates\IEEEauthorrefmark{2},
Huifeng Guo\IEEEauthorrefmark{3},
Ruiming Tang\IEEEauthorrefmark{3},
Xiuqiang He\IEEEauthorrefmark{3} 
}
\IEEEauthorblockA{\IEEEauthorrefmark{1}Montreal Research Center\\
Huawei Noah's Ark Lab,
Montreal, QC, Canada\\ 
\{jianing.sun, yingxue.zhang\}@huawei.com}
\IEEEauthorblockA{\IEEEauthorrefmark{2} McGill University, Montreal, QC, Canada\\
chen.ma2@mail.mcgill.ca, mark.coates@mcgill.ca}
\IEEEauthorblockA{\IEEEauthorrefmark{3}Huawei Noah's Ark Lab, Shenzhen, China\\
\{huifeng.guo, tangruiming, hexiuqiang1\}@huawei.com}}


\maketitle

\begin{abstract}
  Personalized recommendation is ubiquitous, playing an important role
  in many online services. Substantial research has been dedicated to
  learning vector representations of users and items with the goal of
  predicting a user's preference for an item based on the similarity
  of the representations. Techniques range from classic matrix
  factorization to more recent deep learning based
  methods. However, we argue that existing methods do not make full use of the
  information that is available from user-item interaction data and
  the similarities between user pairs and item pairs.
  In this work, we develop a graph convolution-based recommendation
  framework, named Multi-Graph Convolution Collaborative Filtering
  (Multi-GCCF), which explicitly incorporates multiple graphs in the
  embedding learning process. Multi-GCCF not only expressively models
  the high-order information via a bipartite user-item interaction graph, but
  also integrates the proximal information by building and processing
  user-user and item-item graphs. Furthermore, we consider the intrinsic difference between user nodes and item nodes when performing graph convolution on the bipartite graph. We conduct extensive experiments on
  four publicly accessible benchmarks, showing significant
  improvements relative to several state-of-the-art collaborative
  filtering and graph neural network-based recommendation models. Further experiments quantitatively verify the
  effectiveness of each component of our proposed model and demonstrate that
  the learned embeddings capture the important relationship structure.

\end{abstract}


% \fi
\IEEEpeerreviewmaketitle

\section{Introduction}
\label{sec:intro}
Rapid and accurate prediction of users' preferences is the ultimate
goal of today's recommender systems~\cite{Ricci-recsys}. Accurate
personalized recommender systems benefit both demand-side and
supply-side, including the content publisher and platform. Therefore,
recommender systems not only attract great interest in
academia~\cite{fm,CF,ffm}, but also are widely developed in industry
\cite{wide-deep,dnn-youtube}. The core method behind recommender systems is collaborative
filtering (CF)~\cite{cf-rs,mf-rs}. The
basic assumptions underpinning collaborative filtering are that
similar users tend to like the same item and items with
similar audiences tend to receive similar ratings from an individual.

One of the most successful methods for performing collaborative
filtering 
% and hence designing effective recommender systems, 
is matrix factorization (MF)~\cite{mf-rs, pLSA, svd++}. MF models characterize both
items and users by vectors in the same space, inferred from the
observed entries of the user-item historical interaction. More recently, deep learning models have been introduced to boost the performance of traditional MF models. However, as observed in~\cite{NGCF_wang19}, deep learning-based
recommendation models are not sufficient to yield optimal embeddings
because they consider only user and item features.
There is no explicit incorporation of user-item {\em interactions} when
developing embeddings; the interactions are only used to define the
learning objectives for the model training. A second limitation of the deep learning models is the reliance on the
explicit feedback from users, which is usually relatively sparse. 

Bearing these limitations in mind, a natural strategy is to develop mechanisms to directly involve the user-item interactions in the
embedding construction. Recent works by Ying et al.~\cite{ying2018} and Wang et
al.~\cite{NGCF_wang19} have demonstrated the effectiveness of
processing the bipartite graph, reporting improvements over the
state-of-the-art models. 

Despite their effectiveness, we perceive two important limitations. \emph{First}, these models ignore the intrinsic difference between the
two types of nodes in the bipartite graph (users and items). When
aggregating information from neighboring nodes in the graph during
the embedding construction procedure, the architectures
in~\cite{ying2018,NGCF_wang19} combine the information in the same
way, using a function that has no dependence on the nature of the node. However,
there is an important intrinsic difference between users and items in a real environment.
This suggests that the aggregation and transformation functions should be dependent on the type of entity. \emph{Second}, user-user and item-item relationships are also very important signals. Although
two-hop neighborhoods in the bipartite graph capture these to some
extent, it is reasonable to assume that we can improve the
recommendation quality by constructing and learning from graphs that directly model
user-user and item-item relationships.

In this paper, we propose a novel graph convolutional neural network (GCNN)-based recommender system framework, {\em Multi-GCCF}, with two key innovations:

\begin{itemize}[leftmargin=*]
\item
  {\em Capturing the intrinsic difference between users and items}: we apply separate
  aggregation and transformation functions to process user nodes and item nodes when learning with a graph neural network.

  We find that the user and item embeddings are learned more precisely and the recommendation performance is improved.

\item {\em Modeling user-user and item-item relationships explicitly}:
  we construct separate user-user and
  item-item graphs. Multi-GCCF conducts learning simultaneously on all three
  graphs and employs a multi-graph encoding layer to
  integrate the information provided by the user-item, user-user, and
  item-item graphs. 
\end{itemize}

We conduct empirical studies on four real-world datasets, which 
comprise more than one million user-item interactions. Extensive results demonstrate the superiority
of Multi-GCCF over the strongest state-of-the-art models.

\section{Related Work}
\subsection{Model-based Collaborative Filtering methods}

Model-based CF methods learn the similarities between items and users
by fitting a model to the user-item interaction data.  Latent factor
models are common, such as probabilistic Latent Semantic Analysis
(pLAS~\cite{pLSA}) and the most widely used approach, Matrix
Factorization (MF~\cite{mf-rs}). Koren proposed SVD++~\cite{svd++}, which combines information about a user's ``neighbors", i.e. items she has previously interacted with, and matrix factorization for
prediction. Factorization machines~\cite{fm,ffm} provide a mechanism
to incorporate side information such as user demographics and item
attributes. 

MF-based methods are limited because they are confined to the
inner-product as a mechanism for measuring the similarity between
embeddings of users and items. Recently, neural networks have been
incorporated into collaborative filtering architectures~\cite{NCF,fgcnn,xdeepfm, deepfm,pin}. These use a
combination of fully-connected layers, convolution, inner-products and
sub-nets to capture complex similarity relationships.

\begin{figure*}[ht!]
\centering
  \includegraphics[width=0.78\textwidth]{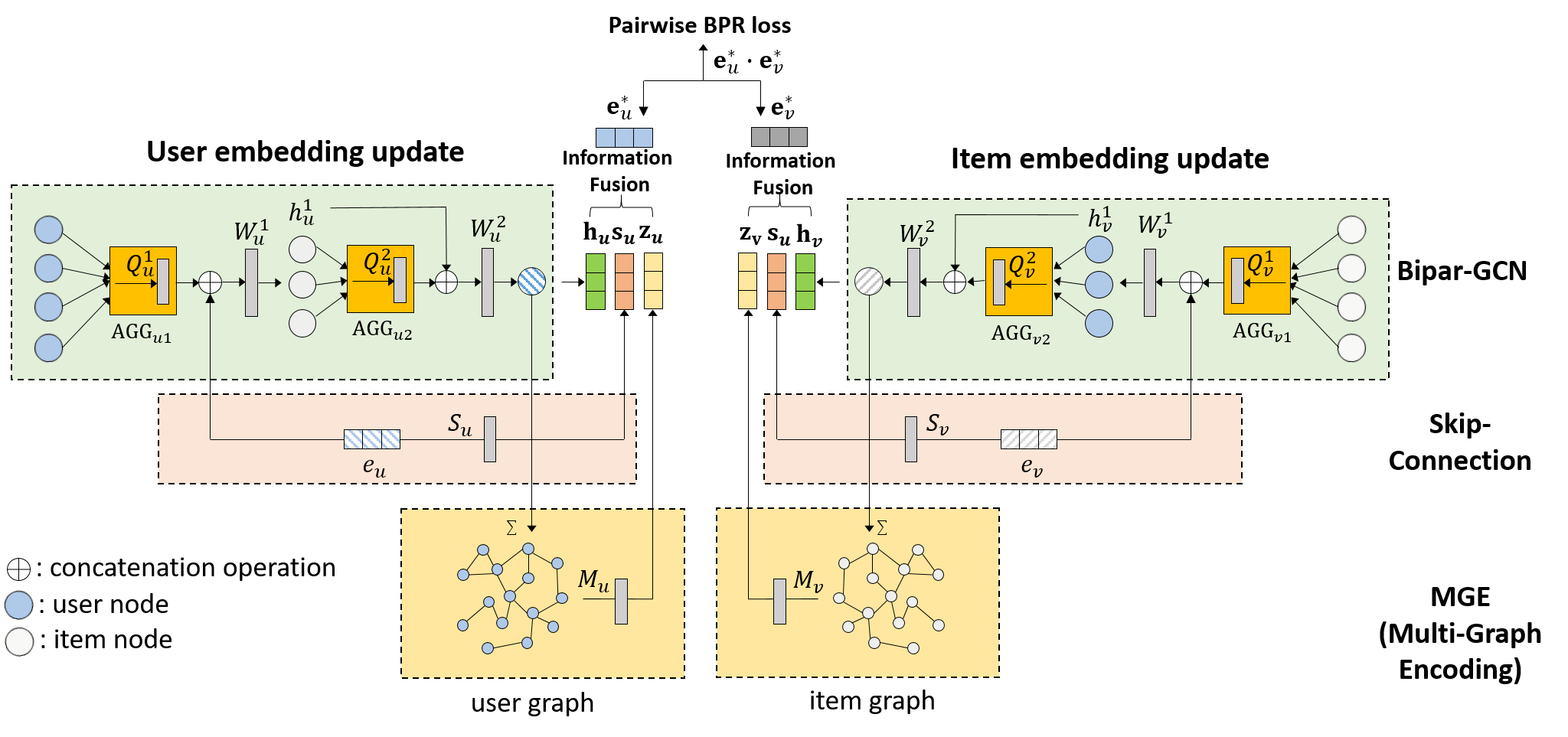}
  \caption{The overall architecture of Multi-GCCF.}
   \label{fig:overall-framework}
\vspace{-0.5cm}
\end{figure*}

\subsection{Graph-based recommendation}
Graphs are a natural tool for representing rich pairwise relationship information in recommendation systems.
Early works~\cite{item-rank,bi-rank,hop-rec} used label propagation
and random walks on the user-item interaction graph to derive
similarity scores for user-item pairs. With the emerging field in Graph Neural Networks (GNNs)~\cite{kipf2017,hamilton2017,zhang2018graph,zhang2019}, more recent works have started
to apply graph neural networks~\cite{gcmc_vdberg2018,ying2018,NGCF_wang19}. Graph Convolutional Matrix Completion
(GCMC)~\cite{gcmc_vdberg2018} treats the recommendation problem as a
matrix completion task and employs a graph convolution autoencoder.
PinSAGE~\cite{ying2018} applies a graph neural network on the item-item
graph formed by modeling the similarity between
items. 
% PinSAGE has been reported to improve significant performance
% gains for the Pinterest recommendation system. 
Neural Graph Collaborative Filtering (NGCF)~\cite{NGCF_wang19} processes the
bipartite user-item interaction graph to learn user
and item embeddings.

\section{Methodology}
\label{sec:meth}

In this section, we explain the three key components of our
method. \emph{First}, we develop a Bipartite Graph Convolutional
Neural Network (Bipar-GCN) that acts as an encoder to generate user
and item embeddings, by processing the user-item interaction bipartite
graph. \emph{Second}, a Multi-Graph Encoding layer (MGE) encodes
latent information by constructing and processing multiple graphs:
besides the user-item bipartite graph, another two graphs represent
user-user similarities and item-item similarities
respectively. \emph{Third}, a skip connection structure between the
initial node feature and final embedding allows us to exploit any
residual information in the raw feature that has not been captured by
the graph processing. The overall framework of Multi-GCCF is depicted
in Figure~\ref{fig:overall-framework}.

\subsection{Bipartite Graph Convolutional Neural Networks}

\begin{figure}[htbp]
    \centering
    \includegraphics[width=0.9\linewidth]{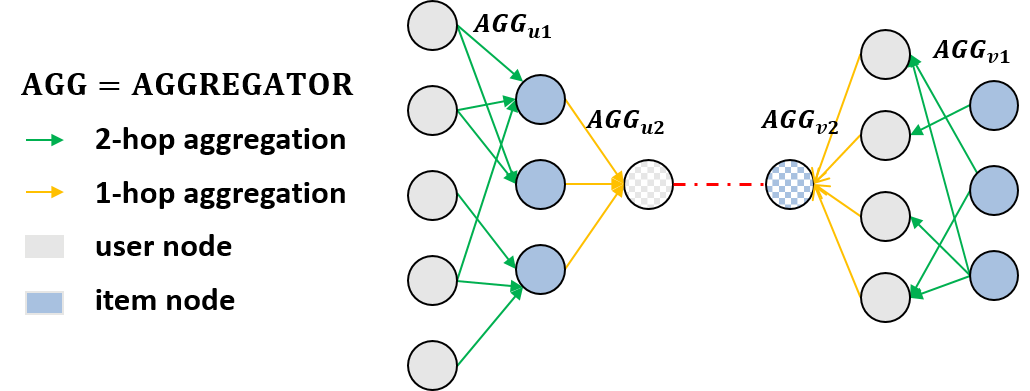}
    \caption{The accumulation of information in the bipartite user-item interaction graph. }
    %   The circles with a mosaic pattern
    %   are target nodes (users on the left, items on the right) that are selected from
    %   the current training batch. Information is fused
    %   from the two-hop neighbours of a node. Only the embeddings of target nodes
    %   are updated during each iteration of the training procedure.}
    \label{fig:bipart_graph}
    \vspace{-0.5cm}
\end{figure}

In a recommendation scenario, the user-item interaction can be readily formulated as a bipartite graph with two types of nodes. 
% Because these two types of nodes serve very different roles and represent different entities, we process them differently and separately when learning in the graph convolution framework.
We apply a Bipartite Graph Convolutional Neural Network (Bipar-GCN) with one side representing user nodes and the other side representing item nodes, as shown in Figure~\ref{fig:bipart_graph}.
The Bipar-GCN layer consists of two phases:~\emph{forward sampling} and~\emph{backward aggregating}. The forward sampling phase is designed to deal with the long-tailed nature of the degree distributions in the bipartite graph. For example, popular items may attract many interactions from users while other items may attract very few. 

After sampling the neighbors from layers $1$ to $K$, Bipar-GCN encodes
the user and item nodes by iteratively aggregating $k$-hop
neighborhood information via graph convolution. There are initial
embeddings $\mathbf{e}_u$ and $\mathbf{e}_v$ that are learned for each
user $u$ and item $v$. These embeddings are learned at the same time
as the parameters of the GCNs. If there are informative input features
$\mathbf{x}_u$ or $\mathbf{x}_v$, then the initial embedding can be a
function of the features (e.g., the output of an MLP applied to
$\mathbf{x}_u$).
% in the absence of informative features, we learn the
% $\mathbf{e}_u$ and $\mathbf{e}_v$ vectors as parameters of the model.
The layer-$k$ embeddings of the target user $u$ can be represented as:
\begin{equation}
 \mathbf{h}^k_{u} = \sigma\Big(\mathbf{W}_u^k\cdot \textbf{[}
 \mathbf{h}_u^{k-1}; \mathbf{h}_{\mathcal{N}(u)} ^{k-1}
 \textbf{]}\Big),
 \hspace{0.2cm}{\mathbf{h}}^0_{u} = {\mathbf{e}}_u \, ,
\end{equation}
where $\mathbf{e}_u$ are the initial user embeddings, $[\,\,;\,]$ represents concatenation, $\sigma(\cdot)$ is the
$\mathrm{tanh}$ activation function, $\mathbf{W}_u^k$ is the
layer-$k$ (user) transformation weight matrix shared across all user nodes. $\mathbf{h}_{\mathcal{N}(u)} ^{k-1}$ is the learned neighborhood embedding. To achieve permutation invariance in the neighborhood, we apply an element-wise weighted mean aggregator:
\begin{align}
 &\mathbf{h}_{\mathcal{N}(u)} ^{k-1} =  \mathrm{AGGREGATOR_u}\Big(
 \left\{\mathbf{h}^{k-1}_{v},  v \in \mathcal{N}(u)\right\} \Big) \,,\\
  &\mathrm{AGGREGATOR_u} = \sigma \left( \mathrm{MEAN}\left( \left\{
    \mathbf{h}_{v} ^{k-1}\cdot \mathbf{Q}_u^k , v \in \mathcal{N}(u) \right\} \right) \right). \nonumber
\end{align}

  Here $\mathbf{Q}_u^k$ is the layer-$k$ (user) aggregator weight
  matrix, which is shared across all user nodes at layer $k$, and
  $\mathrm{MEAN}$ denotes the mean of the vectors in the argument set.

Similarly, the embedding of target item node $v$ can be generated using
another set of (item) transformation and aggregator weight matrices:
\vspace{-0.2cm}

\begin{align}
  & \mathbf{h}^k_{v} = \sigma\Big(\mathbf{W}_v^k\cdot \textbf{[}
 \mathbf{h}_v^{k-1}; \mathbf{h}_{\mathcal{N}(v)} ^{k-1}
 \textbf{]}\Big), \hspace{0.2cm}{\mathbf{h}}^0_{v} = {\mathbf{e}}_v\,, \nonumber \\ 
  &\mathbf{h}_{\mathcal{N}(v)} ^{k-1} = \mathrm{AGGREGATOR_v}\Big(\left\{ \mathbf{h}^{k-1}_{u}, u \in \mathcal{N}(v) \right\} \Big) \,, \\ 
 & \mathrm{AGGREGATOR_v} = \sigma \left( 
    \mathrm{MEAN}\left( \left\{
         \mathbf{h}_{u} ^{k-1} \cdot \mathbf{Q}_v^k,  u \in \mathcal{N}(v)
        \right\} \right) \right) \,. \nonumber 
\end{align}

% \begin{equation}
%   \mathrm{AGGREGATOR_v} = \sigma \left( 
%     \mathrm{MEAN}\left( \left\{
%          \mathbf{h}_{u} ^{k-1} \cdot \mathbf{Q}_v^k,  u \in \mathcal{N}(v)
%         \right\} \right) \right) \,
% \end{equation}

\subsection{Multi-Graph Encoding Layer} 

% Collaborative filtering models suffer from a data sparsity problem, which can
% severely limit the models' capacity.
% Furthermore, the second-order
% proximity relationships (two-hop neighbors) in the bipartite graph representing
% user-item interaction can \emph{only implicitly} represent user-user and item-item relationship information. This information could be more useful for recommendation performance if represented \emph{explicitly}.
To alleviate the data sparsity problem in CF, we propose a Multi-Graph Encoding (MGE)
layer, which generates an additional embedding for a target user or
item node by constructing two additional graphs and applying graph
convolutional learning on them. 
% The multi-graph neural network
% learning task has been investigated in~\cite{monti2017geometric,wu2019wechat, geng2019spatiotemporal}. In these works, multiple graphs are constructed to reflect different dependency relationships. GCNN learning is applied to each of the
% different graphs before a final merged output is derived by applying
% an information aggregation function such as concatenation,
% summation~\cite{geng2019spatiotemporal} or a multi-arm bandit
% (MAB)~\cite{wu2019wechat}.

In particular, in addition to the user-item bipartite graph, we
construct a user-user graph and an item-item graph to capture the
proximity information among users and items. This proximity
information can make up for the very sparse user-item interaction
bipartite graph. 
% If there is no side-information about users or items, then 
The graphs are constructed by computing pairwise cosine
similarities on the rows or columns of the rating/click matrix.
% We apply a
% threshold to determine if there is an edge or not in the constructed graph.

% If there is side information such as a social graph (user-user graph)
% or geographical location (item-item graph based on the latitude and
% longitude), then we can process the provided graphs directly or
% construct graphs using more general similarity functions.

In the MGE layer, we generate embeddings for target nodes by aggregating
the neighborhood features using a one-hop graph convolution
layer and a sum aggregator:
\begin{equation}
\scalemath{0.85}{
\begin{split}
    \mathbf{z}_u = \sigma\Big(\sum_{i\in\mathcal{N}'(u)} \mathbf{e}_i \cdot \mathbf{M}_u\Big)  \,;\quad \mathbf{z}_v = \sigma\Big(\sum_{j\in\mathcal{N}'(v)} \mathbf{e}_j \cdot \mathbf{M}_v\Big)  \,.
\end{split}}
\end{equation}
Here $\mathcal{N}'(u)$ denotes the one-hop neighbourhood of user $u$
in the user-user graph and $\mathcal{N}'(v)$ denotes the one-hop
neighbourhood of item $v$ in the item-item graph.  $\mathbf{M}_u$ and
$\mathbf{M}_v$ are the learnable user and item aggregation weight
matrices, respectively.

In contrast to the Bipar-GCN layer, no additional neighbour sampling
is performed in the MGE layer. 
% We aggregate over all direct neighbors
% in the user-user and item-item graphs because the constructed graphs
% do not have a long-tailed degree distribution. 
We select thresholds
based on the cosine similarity that lead to an average degree of 10 for
each graph.

% The embeddings $\mathbf{z}_u$ and $\mathbf{z}_v$ contain different
% information from the learned embeddings $\mathbf{h}_u$ and
% $\mathbf{h}_v$ from the Bipar-GCN layer. 
By merging the outputs of the Bipar-GCN and MGE layers together, we can
take advantage of the different dependency relationships encoded
by the three graphs. All three graphs can be easily constructed from
historical interaction data alone, with very limited additional
computation cost.

% Even in the case where there is no additional content information and
% all graphs are constructed based on the same user-item interaction
% data (for example, clicking or rating history), the embeddings derived
% from them are acquired from different perspectives, leading to more
% informative and comprehensive latent representations. We can also view
% the graphs from the perspective of regularization; each graph enforces
% that the learned embedding is smooth with respect to the graph
% connectivity~\cite{zhu2003semi,
%   zhou2004semilearning}. The use of multiple graphs can impose
% additional regularization and alleviate
% performance bottlenecks related to overfitting and the cold start problem.

%\vspace{-0.1cm}
\subsection{Skip-connection with Original Node Features}

We further refine the embedding with information passed directly from the
original node features. The intuition behind this is that both Bipar-GCN and MGE focus on extracting latent information based on relationships. As a result, the impact of the initial node features becomes less dominant. The skip connections allows the architecture to re-emphasize these features.

We pass the original features through a single fully-connected layer
to generate skip-connection embeddings.

\subsection{Information Fusion}

The bipartite-GCN, MGE layer and skip connections reveal latent information from three perspectives. 
%First, the bipartite-GCN captures behavioural similarity between users and items by explicitly modeling the historical interaction as a bipartite graph. Second, the MGE layer identifies similarities among users and items by constructing user-user and item-item graphs. Third, the skip connections allow learning from individual node characteristics by re-emphasizing initial features.
It is important to determine how to merge these different embeddings
effectively. In this work we investigate three
methods to summarize the individual embeddings into a single embedding
vector: ~\emph{element-wise sum}, ~\emph{concatenate},
and~\emph{attention mechanism}. The exact operation of these three
methods is described in Table~\ref{tab2}. We experimentally compare them in Section~\ref{sec:exp}.

\begin{table}[htbp]
\caption{Comparison of different message fusion methods.}
    \centering
\resizebox{\linewidth}{!}{
\Huge
\begin{tabular}{c|c}
\hline\hline
\multirow{2}{*}{} & Formula \\ \hline
Element-wise sum & $\mathbf{e}^*_u = \mathbf{h}_u^K + \mathbf{z}_u + \mathbf{s}_u$ \\ \hline
Concatenation & $\mathbf{e}^*_u = [\mathbf{h}_u^K ; \mathbf{z}_u ; \mathbf{s}_u]$  \\ \hline
\multirow{2}{*}{Attention} & 
$\mathbf{A}_u = Softmax \Big(W_{a_s} \cdot \sigma\big(W_{a_1}\cdot \mathbf{h}^K_{u} + W_{a_2}\cdot \mathbf{z}_u + W_{a_3} \cdot \mathbf{s}_u\big)\Big)$ \\ 
& $\mathbf{e}^*_u = \textbf{\big[}\mathbf{h}^K_{u}; \mathbf{z}_u; \mathbf{s}_u\textbf{\big]}\cdot \mathbf{A}_u $ \\ \hline\hline
\end{tabular}}
\vspace{-0.5cm}
\label{tab2}
\end{table}

\subsection{Model Training}
We adapt our model to allow forward and backward propagation for
mini-batches of triplet pairs $\{u, i, j\}$. To be more specific, we
select unique user and item nodes $u$ and $v=\{i, j\}$ from mini-batch
pairs, then obtain low-dimensional embeddings
$\{\mathbf{e}_u, \mathbf{e}_i, \mathbf{e}_j\}$ after information fusion,
with stochastic gradient descent on the widely-used Bayesian
Personalized Recommendation (BPR)~\cite{RendleFGS2009_bpr} loss for
optimizing recommendation models. The objective function is as
follows:
\begin{equation}
\scalemath{0.9}{
\begin{split}
loss = \sum_{(u, i, j)\in\mathcal{O}}&-\log\sigma( \mathbf{e}^*_u \cdot \mathbf{e}^*_i - \mathbf{e}^*_u \cdot \mathbf{e}^*_j) \\ 
& + \lambda||\Theta||_2^2 + \beta(||\mathbf{e}^*_u||_2^2+||\mathbf{e}^*_i||_2^2+||\mathbf{e}^*_j||_2^2) \,
\end{split}
}
\end{equation}
where
$\mathcal{O}=\{(u,i,j)|(u,i)\in\mathcal{R^+},
(u,j)\in\mathcal{R^-})\}$ denotes the training batch. $\mathcal{R^+}$
indicates observed positive interactions. $\mathcal{R^-}$ indicates
sampled unobserved negative interactions. $\Theta$ is the model parameter set and $\mathbf{e}^*_u$, $\mathbf{e}^*_i$, and $\mathbf{e}^*_j$  are the learned embeddings. We conduct regularization on both model parameters and generated embeddings to prevent overfitting (regularization coefficients $\lambda$ and $\beta$).

% GNNs have also been employed for the task of social recommendation
% in~\cite{fan2019graph,song2019session_dynamic,wu2019wechat} and the
% incorporation of knowledge graph information in news recommendation~\cite{wang2018dkn}. 

% As discussed in Section~\ref{sec:intro}, our proposed model is most
% closely related to the NGCF~\cite{NGCF_wang19} and
% PinSage~\cite{ying2018} collaborative filtering methods. The key
% differences are that we learn over multiple graphs (user-item,
% user-user, and item-item) and that we employ node-dependent
% information aggregation functions, so that we can fuse information
% differently depending on whether it is being collected from nodes
% representing users or items. 

% \textbf{Neighborhood Dropout.}~Deep learning often suffers from
% overfitting when the model becomes complicated and the number of
% parameters increases. For graph neural networks, node dropout is an
% effective solution to not only prevent overfitting but also boost the
% model's ability to generalize to unobserved
% interactions~\cite{gcmc_vdberg2018}. Instead of dropping out some
% nodes completely with some probability, we apply message dropout on
% the aggregated neighborhood features for each target node, making
% embeddings more robust against the presence or absence of single
% edges.

\begin{table*}[t]
\caption{The overall performance comparison. Underline indicates the second best model performance. Asterisks denote scenarios where a Wilcoxon signed rank test indicates a statistically significant difference between the scores of the best
and second-best algorithms. }
\vspace{-0.2cm}
    \centering
\resizebox{\linewidth}{!}{
\small
\begin{tabular}{c|c|c|c|c|c|c|c|c}
\hline\hline
\multirow{2}{*}{} &
\multicolumn{2}{c|}{Gowalla} & %
    \multicolumn{2}{c|}{Amazon-Books} &
    \multicolumn{2}{c|}{Amazon-CDs} &
  %   \multicolumn{2}{c|}{Movielens-10M} &
    \multicolumn{2}{c}{Yelp2018}\\
\cline{2-9}
 & Recall@20 & NDCG@20 & Recall@20 & NDCG@20 &Recall@20 & NDCG@20 &Recall@20 & NDCG@20 \\ \hline
BPRMF & 0.1291 & 0.1878 & 0.0250 & 0.0518 & 0.0865 & 0.0849 & 0.0494 &0.0662 \\ 
NeuMF & 0.1326 & 0.1985 & 0.0253 & 0.0535 & 0.0913 & 0.1043 & 0.0513 & 0.0719 \\ \hline
GC-MC & 0.1395 &  0.1960 & 0.0288 & 0.0551 & 0.1245 & 0.1158 & 0.0597 & 0.0741 \\ 
PinSage & 0.1380 & 0.1947 & 0.0283 & 0.0545 & 0.1236 & 0.1118 & \underline{0.0612} & \underline{0.0750} \\ 
NGCF & \underline{0.1547} & \textbf{0.2237} & \underline{0.0344} & \underline{0.0630} & \underline{0.1239} & \underline{0.1138} & 0.0581 & 0.0719 \\ \hline
\textbf{Multi-GCCF ($d$=64)} & \textbf{$^*$0.1595} & $^*$0.2126 & \textbf{$^*$0.0363} & \textbf{$^*$0.0656} & \textbf{$^*$0.1390} & \textbf{$^*$0.1271} & \textbf{$^*$0.0667} & \textbf{$^*$0.0810}
\\ 
%\% Improv. & 3.10\% & - & 5.52\% & 4.07\% & 12.19\% & 11.69\% & 9.01\% & 7.95\% \\ \hline
% p-value  & & & & & & & \\ \hline
\textbf{Multi-GCCF ($d$=128)} & \textbf{$^*$0.1649} & $^*$0.2208 & \textbf{$^*$0.0391} & \textbf{$^*$0.0705} &  \textbf{$^*$0.1543} & \textbf{$^*$0.1350} &\textbf{$^*$0.0686} & \textbf{$^*$0.0835} \\ \hline \hline 
%\% Improv. & 6.39\% & - & 13.66\% & 11.90\% & 24.54\% & 23.89\% & 12.41\% & 10.67\% \\ \hline \hline
\end{tabular}}
\vspace{-0.3cm}
\label{results}
\end{table*}

\section{Experimental Evaluation}
\label{sec:exp}
We perform experiments on four real-world datasets to evaluate our
model. Further, we conduct extensive ablation studies on each proposed component (Bipar-GCN, MGE and skip connect). We also provide a visualization of the learned representation.

\vspace{-0.2cm}
\subsection{Datasets and Evaluation Metrics}
To evaluate the effectiveness of our method, we conduct extensive experiments on four benchmark datasets:~\emph{Gowalla},~\emph{Amazon-Books, Amazon-CDs} and~\emph{Yelp2018 \footnote{https://snap.stanford.edu/data/loc-gowalla.html; http://jmcauley.ucsd.edu/data/amazon/; https://www.yelp.com/dataset/challenge}}. These datasets are publicly accessible, real-world data with various domains, sizes, and sparsity. For all datasets, we filter out users and items with fewer than 10 interactions. Table~\ref{tab1} summarizes their statistics.

For all experiments, we evaluate our model and baselines in terms of
\textit{Recall@k} and \textit{NDCG@k} (we report \textit{Recall@20} and \textit{NDCG@20}). \textit{Recall@k} indicates the coverage of true (preferred) items as a result of top-$k$ recommendation. \textit{NDCG@k} (normalized discounted cumulative gain) is a measure of ranking quality.
% \begin{itemize}[leftmargin=*]
%     \item \textit{Recall@k} indicates the coverage of
%       true (preferred) items as a result of top-$k$ recommendation. The value is
%       computed by $\mathrm{Recall@k}=\frac{|\mathcal{I}_u^+\cap
%         I_k(u)|}{|\mathcal{I}_u^+|}$, where the target user
%       $u\in\mathcal{U}$, $\mathcal{I}$ is the set of all items,
%       $I_k(u)\in\mathcal{I}$ is an ordered set of the top-$k$ items,
%       $\mathcal{I}_u^+$ is the set of true (preferred) items, and
%       $|\mathcal{I}_u^+\cap I_k(u)|$ is the number of true positives.
      
%     \item \textit{NDCG@k}: The Normalized Discounted Cumulative Gain (NDCG)~\cite{NGCF_wang19}
%       computes a score for $I(u)$ which emphasizes
%       higher-ranked true positives. It is
%       computed as
%       $$\mathrm{NDCG}@k=\frac{\mathrm{DCG_k}}{\mathrm{IDCG_k}}=\frac{\sum_{n=1}^{|\mathcal{I}|}\mathrm{D}_k(n)[i_n\in\mathcal{I}_u^+]}{\sum_{n=1}^{|I_k(u)|}D_k(n)}$$ 
%       where $D_k(n)=(2^{rel_{n}}-1)/\log_2(n+1)$ for a relevancy score
%       $rel_n$. We only consider
%       binary responses, so we use a binary relevance score:
%       $\mathrm{rel}_n=1$ if $i_n\in\mathcal{I}_u^+$ and 0 otherwise.
% \end{itemize}

\begin{table}[htbp]
\caption{Statistics of evaluation datasets.}
\vspace{-0.3cm}
\begin{center}
\begin{tabular}{c|c|c|c|c}
\hline
Dataset & \#User & \#Items & \#Interactions & Density \\ \hline\hline
Gowalla & 29,858 & 40,981 & 1,027,370 & 0.084\% \\ \hline
Yelp2018 & 45,919 & 45,538 & 1,185,065 & 0.056\% \\ \hline
Amazon-Books & 52,643 & 91,599 & 2,984,108 & 0.062\% \\ \hline
% Movielens-10M & 69,878 & 10,677 & 10,000,054 & 0.134\% \\ \hline
Amazon-CD & 43,169 & 35,648 & 777,426 & 0.051\% \\ \hline
\end{tabular}
\label{tab1}
\end{center}
\vspace{-0.5cm}
\end{table}

\subsection{Baseline Algorithms}
We studied the performance of the following models.\\ 
{Classical collaborative filtering methods}: \textbf{BPRMF}~\cite{RendleFGS2009_bpr}.  \textbf{NeuMF}~\cite{NCF}.  Graph neural network-based collaborative filtering methods: \textbf{GC-MC}\cite{gcmc_vdberg2018}.  \textbf{PinSage}~\cite{pin}.  \textbf{NGCF}~\cite{NGCF_wang19}. 

Our proposed method: \textbf{Multi-GCCF}, which contains two graph convolution layers on the user-item bipartite graph (2-hop aggregation), and one graph convolution layer on top of both the user-user graph and the item-item graph to model the similarities between user-pairs and item-pairs.

\subsection{Parameter Settings}
We optimize all models using the Adam optimizer with the xavier initialization. The embedding
size is fixed to 64 and the batch size to 1024, for all baseline
models. Grid search is applied to choose the learning rate and the
coefficient of $L_2$ normalization over the ranges \{0.0001, 0.001,
0.01, 0.1\} and \{$10^{-5}$, $10^{-4}$, \dots, $10^{-1}$\},
respectively. As in~\cite{NGCF_wang19}, for GC-MC and NGCF, we also
tune the dropout rate and network structure. Pre-training~\cite{NCF}
is used in NGCF and GC-MC to improve performance. We implement our Multi-GCCF model in PyTorch and use two Bipar-GCN
layers with neighborhood sampling sizes $S_1=15$ and $S_2=10$. The
output dimension of the first layer is fixed to 128; the final output
dimension is selected from $\{64, 128\}$ for different experiments. We set the input node embedding dimension to $512$. The
neighborhood dropout ratio is set to $0.2$. The regularization
parameters in the objective function are set to $\lambda=0.01$ and
$\beta=0.02$.

\subsection{Comparison with Baselines}
Table~\ref{results} reports the overall performance compared with
baselines. Each result is the average performance from 5 runs with
random weight initializations.

We make the following observations:
\begin{itemize}[leftmargin=*]
\item Multi-GCCF consistently yields the best performance for all datasets. More precisely, Multi-GCCF improves over the strongest baselines with respect to recall@20 by $9.01$\%, $12.19$\%, $5.52$\%, and $3.10$\% for Yelp2018, Amazon-CDs, Amazon-Books and Gowalla, respectively. Multi-GCCF further outperforms the strongest baselines by $12.41$\%, $15.43$\%, $24.54$\% and $6.39$\% on recall@20 for Yelp2018, Amazon-CDs, Amazon-Books and Gowalla, respectively, when increasing the latent dimension.
For the NDCG@20 metric, Multi-GCCF outperforms the next best method by $5$\% to $25$\% on three dataset. Further,  This suggests that, exploiting the latent information by utilizing multiple graphs and efficiently integrating different embeddings, Multi-GCCF ranks relevant items higher in the recommendation list.

  %\item By replacing the inner product with a multi-layer perceptron, NeuMF surpasses BPRMF in all cases. However, as a result of failing to model the connectivity of users and items, NeuMF performs worse than GC-MC and PinSage.

\end{itemize}

\begin{table}[htbp]
\caption{Ablation studies.}
\vspace{-0.2cm}
    \centering
\resizebox{0.8\linewidth}{!}{
\Huge
\begin{tabular}{c|c|c}
\hline\hline
\multirow{2}{*}{Architecture} & \multicolumn{2}{c}{Yelp2018}\\
\cline{2-3} 
 & Recall@20 & NDCG@20 \\ \hline
Best baseline ($d$=64)  & 0.0612 & 0.0744 \\ \hline 
Best baseline ($d$=128) & 0.0527 & 0.0641 \\ \hline 
1-hop Bipar-GCN & 0.0650 & 0.0791 \\ \hline
2-hop Bipar-GCN & 0.0661 & 0.0804 \\ \hline
2-hop Bipar-GCN + skip connect & 0.0675 & 0.0821  \\ \hline
2-hop Bipar-GCN + MGE & 0.0672 & 0.0818  \\ \hline
\textbf{Multi-GCCF ($d$=128)} & \textbf{0.0686} & \textbf{0.0835}\\ \hline \hline
\end{tabular}}

\label{ablation}
\end{table}

\begin{figure}[t]
\vspace{-0.2cm}
\centering
\begin{subfigure}{.5\linewidth}
  \centering
  \includegraphics[width=.9\linewidth]{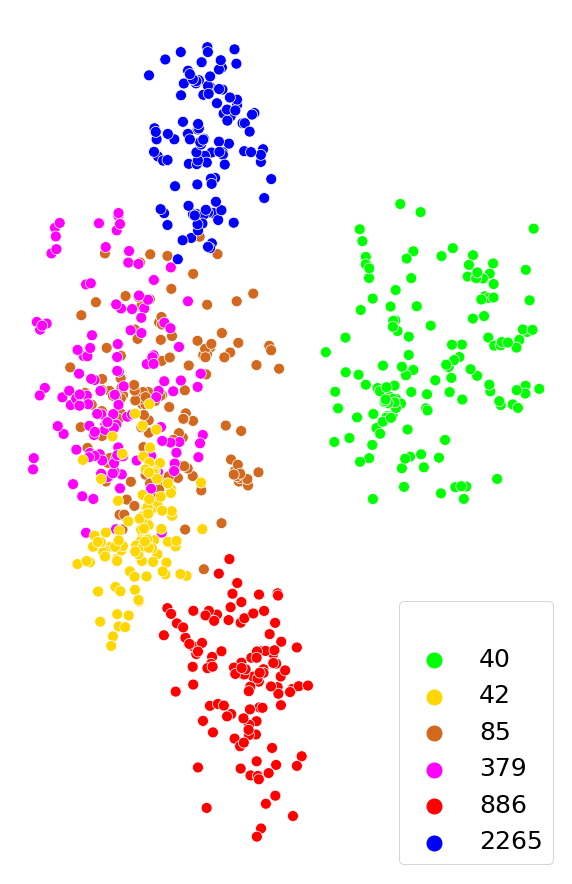}
  \caption{BPRMF}
  \label{fig:ab1}
\end{subfigure}%
\begin{subfigure}{.5\linewidth}
  \centering
  \includegraphics[width=.9\linewidth]{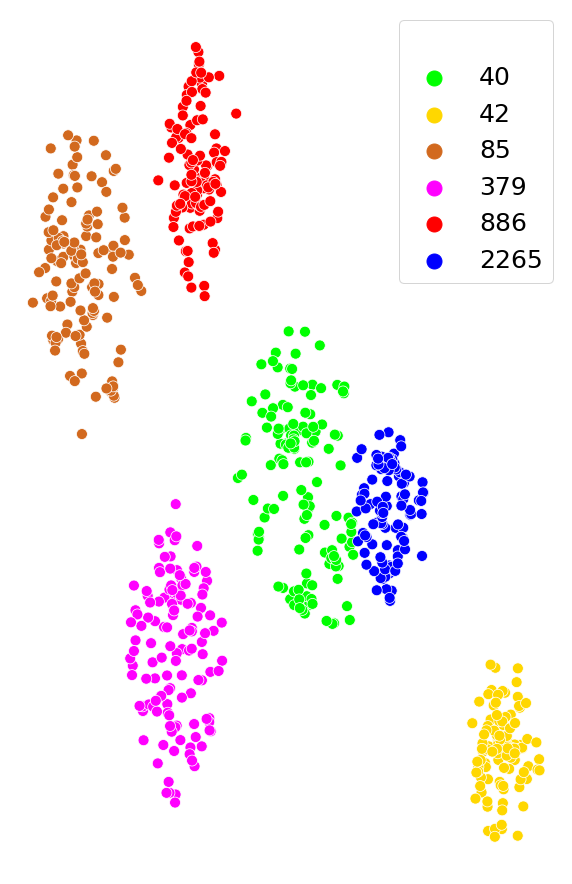}
  \caption{Multi-GCCF}
  \label{fig:ab2}
\end{subfigure}
\caption{Visualization of the t-SNE transformed representations derived between BPRMF and Multi-GCCF on Amazon-CDs. Numbers in the legend are user IDs.}
% Points with the same color represent the relevant items from the corresponding user
\vspace{-0.2cm}
\label{fig:visualize}
\end{figure}

\subsection{Ablation Analysis}
\label{sec:abl}
To assess and verify the effectiveness of the individual components of
our proposed Multi-GCCF model, we conduct an ablation analysis on Gowalla and Yelp2018 in Table~\ref{ablation}. The table illustrates the performance contribution of each component. The output embedding size is 128 for all ablation
experiments. We compare to $d=64$ baselines because they
outperform the $d=128$ versions.

We make the following observations: 
\begin{itemize}[leftmargin=*]
\item All three main components of our proposed model, Bipar-GCN layer, MGE layer, and skip connection, are demonstrated to be effective.
\item Our designed Bipar-GCN can greatly boost the performance
  with even one graph convolution layer on both the user side and the item
  side. Increasing the number of graph convolution layers can slightly improve the
  performance.
\item Both MGE layer and skip connections
  lead to significant performance improvement.
\item Combining all three components leads to further improvement,
  indicating that the different embeddings are effectively capturing different information about users, items, and user-item relationships. 
\end{itemize}

\vspace{-0.05cm}
\begin{table}[htbp]
\caption{Comparison of different information fusion methods when $ d=128 $.}
\vspace{-0.2cm}
    \centering
\resizebox{\linewidth}{!}{
\small
\begin{tabular}{c|c|c|c|c}
\hline\hline
\multirow{2}{*}{} &
\multicolumn{2}{c|}{Gowalla} & %
    \multicolumn{2}{c}{Amazon-CDs}\\
\cline{2-5}
& Recall@20 & NDCG@20 & Recall@20 & NDCG@20 \\ \hline
\textbf{element-wise sum} & \textbf{0.1649} & \textbf{0.2208} & \textbf{0.1543} & \textbf{0.1350} \\ \hline
concatenation & 0.1575 & 0.2179 & 0.1432 & 0.1253 \\ \hline
attention & 0.1615 & 0.2162 & 0.1426 & 0.1248 \\ \hline\hline
\end{tabular}}
\vspace{-0.5cm}
\label{tab22}
\end{table}

\subsection{Effect of Different Information Fusion Methods} 
As we obtain
three embeddings from different perspectives, we compare different
methods to summarize them into one vector: element-wise sum,
concatenation, and attention. Table~\ref{tab22} shows the
experimental results for Gowalla and Amazon-CDs.
We make the following observations: Summation performs much better than concatenation and
  attention. Summation generates an embedding of the same dimension as
  the component embeddings and does not involve any additional
  learnable parameters. The additional flexibility of attention and
  concatenation may harm the generalization capability of the model.  

%\item The sequence provided to the attention mechanism is very short (consisting of only three vectors) and this may inhibit any benefit that attention can provide. In general, attention has been proven to be most effective for long sequences.

\subsection{Embedding Visualization} Figure~\ref{fig:visualize}
provides a visualization of the representations derived from BPRMF
and Multi-GCCF. Nodes with the same color represent all
the item embeddings from one user's clicked/visited history, including
test items that remain unobserved during training.
We find that both BPRMF and our proposed model have the tendency to encode the items that are preferred by the same user close to one another. However, Multi-GCCF generates tighter clusters, achieving a strong grouping effect for items that have been preferred by the same user.

\vspace{-0.2cm}
\section{Conclusion}
In this paper we have presented a novel collaborative filtering
procedure that incorporates multiple graphs to explicitly represent
user-item, user-user and item-item relationships. The proposed model, Multi-GCCF, constructs three embeddings learned from different perspectives on the available data.  Extensive experiments on four real-world datasets demonstrate the effectiveness of our approach, and an ablation study quantitatively verifies that each component makes an important contribution. Our proposed Multi-GCCF approach is well supported under the user-friendly and efficient GNN library developed in MindSpore, a unified training and inference Huawei AI framework. 
%in Huawei's full-stack, all-scenario AI portfolio.

% \section*{Acknowledgment}

% more thanks here



%\bibliographystyle{IEEEtran}


% \end{thebibliography}

% \def\IEEEbibitemsep{0pt plus .8pt}

\bibliographystyle{IEEEtran.bst}
\bibliography{icdm}

\end{document}